# New Heuristics for Interfacing Human Motor System using Brain Waves

Mohammed El-Dosuky[1], Ahmed EL-Bassiouny[2], Taher Hamza[3] and Magdy Rashad[4]

[1] Computer sciences Department, Faculty of Computers and Information,
Mansoura University, Egypt
*mouh_sal_010@mans.edu.eg*

[2] Department of Mathematics, Faculty of Sciences
Mansoura University, Egypt
*el_bassiouny@mans.edu.eg*

[3] Computer sciences Department, Faculty of Computers and Information,
Mansoura University, Egypt
*Taher_Hamza@yahoo.com*

[4] Computer sciences Department, Faculty of Computers and Information,
Mansoura University, Egypt
*magdi_12003@yahoo.com*

**Abstract**
There are many new forms of interfacing human users to machines. We persevere here electric-mechanical form of interaction between human and machine. The emergence of brain–computer interface allows mind-to-movement systems. The story of the Pied Piper inspired us to devise some new heuristics for interfacing human motor system using brain waves, by combining head helmet and LumbarMotionMonitor. For the simulation we use java GridGain. Brain responses of classified subjects during training indicates that Probe can be the best stimulus to rely on in distinguishing between knowledgeable and not knowledgeable.
***Keywords:*** *Pied Piper, Human Motor system, Brain Waves, Brain Fingerprinting, brain–computer interface*

## 1. Introduction

Human-computer interaction (HCI) is a discipline concerned with the design, implementation and evaluation of interactive computing systems for human use and with the study of major phenomenon surrounding them [1].

There are many ways to interface human users to machines. We persevere here electric-mechanical form of interaction between human and machine. This form of interaction utilizes the Electro-Magnetic signals which can be easily measured using EEG or MRI scanning, relating to the well-known surface electromyography for motor system [2].

The Pied Piper story inspired us to devise new heuristics for interfacing human motor system using brain waves. Section 2 reviews previous work. Section 3 reviews the story of the Pied Piper, then introduces the proposed framework. Section 4 evaluates the proposed framework. Section 5 concludes the paper.

## 2. Previous Work

L. Farwell work is centralized at building computerized system for Brain Fingerprinting ([3], [4], [5], [6]) . It is like Polygraph (Lie Detector), but it reads the memory centers of the human brain using multifaceted electroencephalographic response analysis (MERA) instead of measuring physiologic responses such as heart rate, sweating, breathing. This can be used to identify persons thoughts and interpret these thoughts into coordination.

This allows the emergence of brain–computer interface (BCI) as a direct brain-computer communication for assisting, augmenting, or repairing human cognitive or sensory-motor functions, usually based on electroencephalography (EEG) in the following frequency bands ([7], [8], [9]).

Table 1 Frequency bands of the EEG :

| Band | Frequency (Hz) |
|---|---|
| Alpha | 8-12 |
| μ-rhythm | 9-11 |
| Beta | 14 -30 |
| Theta | 4-7 |
| Delta | <3 |

Studies on the operant conditioning empirically proved that monkeys can be taught to control the deflection of a biofeedback meter arm with neural activity[10]. There has been fast progress in BCIs since then[11].

Rhesus monkeys can be trained to use a BCI to follow visual targets on a computer screen with or without assistance of a joystick[12]. A BCI for three-dimensional tracking in virtual reality is created and reproduced control in a robotic arm[13], showing that a monkey can control a robotic arm by its own brain signals to feed itself [14].

BCIs focusing on motor neuro-prosthetics aim to either bring back movement in individuals with handicap or provide devices to assist them, such as interfaces with computers or robot arms [15].

One of the successful stories is BrainGate, a brain implant system developed by the bio-tech company Cyberkinetics in conjunction with the Department of Neuroscience at Brown University [16]. The implanted computer chip converts the thoughts of the user into computer commands allowing to move the cursor to control the computer, just as a mouse is used, enabling him to open emails, and to operate devices such as a television. Recently, two paralyzed people by brainstem stroke are able to control robotic arms for getting and collecting objects [17].

RIKEN BSI-TOYOTA Collaboration Center announced a real-time control of wheelchairs with brain waves [18]. Commands for smooth left and right turns and forward motion of the wheelchair are processed every 125 milliseconds by analyzing brain waves using signal processing technology. Brain-wave analysis data are displayed on a screen in real time, giving neuro-feedback to the driver for efficient operation.

Implantable BCI devices being designed for human use have some issues such as neural signal extraction primarily for motor commands, signal insertion to restore sensation, and technological challenges that remain [19].

A study was conducted to examine the intra- and inter-tester reproducibility of the LumbarMotionMonitor (LMM) as a measure of thoracolumbar range of motion (RoM), velocity and acceleration [20]. Concluding that, even if LMM is a promising device from a clinical and research perspective, the reproducibility of the LMM is suitably high for RoM and velocity for the device to be used for evaluation in a clinical and research setting.

Recently, a framework is proposed for ensuring the accuracy of a three-dimensional LMM for recording dynamic trunk motion characteristics [21].

## 4. Proposed Framework

The story of the Pied Piper tells that the Mayor announced that anybody who could stop rats rampaging around the city, would deserve a reward.

The Pied Piper, played his flute and took the rats into the river and the rats died. The mayor refused to give him the reward, and the community agreed. The Pied Piper took the children away to the mountain. Even a lame child was cured and could walk again.

Let us extract some rules.

- There is only one traveling Pied Piper (*pp*)

- *pp* has a flute (*f*) that can force anything to follow

- Each community has a leading Mayor (*m*).

- The Mayor can announce calls-for-rescue

- The Mayor can give or refuse to give REWARD

- Rats are rampaging around the city

- *pp* has the ability to cure lame child to walk again.

Sensors are spread on the helmet on the patient head as in [18] and on the LMM as in [21]. Each sensor can be signaled as a RAT or a KID. A server is required to play the role of the Mayor. The Pied Piper is an external monitoring unit for assessing, augmenting, and repairing human cognitive or sensory-motor functions.

Preprocessing includes removing unnecessary frequency bands, averaging the current brain activity level, transforming the measured scalp potentials to cortex potentials and denoising. The MERA is then measured. If the total number of sensors response is above a certain threshold, this is called REWARD signal. If there is no REWARD, the Pied Piper interfere and override the operation of the mayor.

For the simulation we use java GridGain [22]. GridGain is an open source grid computing framework that focuses solely on providing the capabilities of adding parallel processing. GridGain 2.0 release incorporates number of new and enhanced features such as redundant mapping, asynchronous and partial reducing, distributed task session supporting "connected" jobs, per-job and per-task distributed resources, annotation-based grid enabling, early and late load balancing capabilities allowing grid task to fully adapt to non-deterministic nature of execution on the grid, universal and dynamic data partitioning and processing co-location supporting most of distributed data caching solutions, native out-of-the-box support for data caching products, and improved runtime metrics, monitoring and management.

The general technical architecture is shown in figure 1:

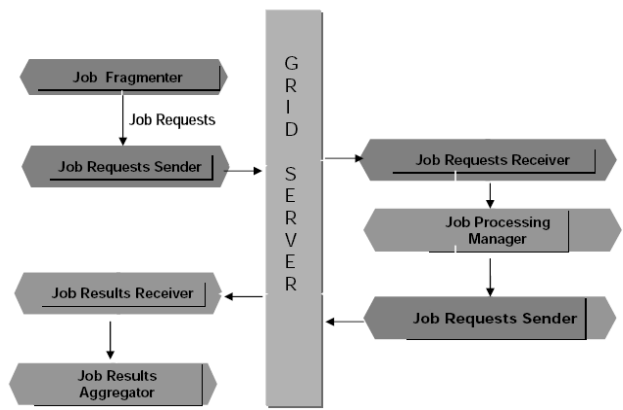

Figure 1 technical Architecture

In a Java Grid there is only one Server that receives work requests from clients and forwards the requests to the workers. It collects work results from the workers and sends back the results to the clients. The Server receives any file required for completing the work request from the client and stores it. The server handles all communication between nodes and any administrative/management task of the grid. There can be as many workers. The worker fulfils work request received by the server and send back the work results. In a Java Grid there can be as many clients as needed. The clients split the work in several work requests and collects work results.

However, the system works in a loop. The general architecture of control unit is shown in the next figure [23].

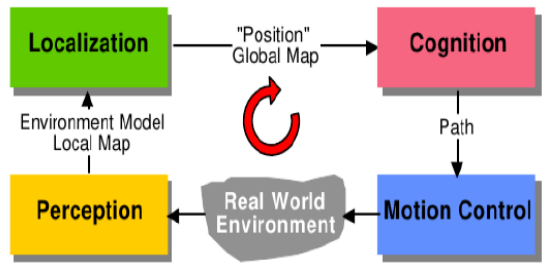

Figure 2 Control Architecture

## 4. Results

Average brain responses to 3 stimuli for 2 subjects classified as knowledgeable and not knowledgeable are shown in figure 3 and figure 4 respectively.

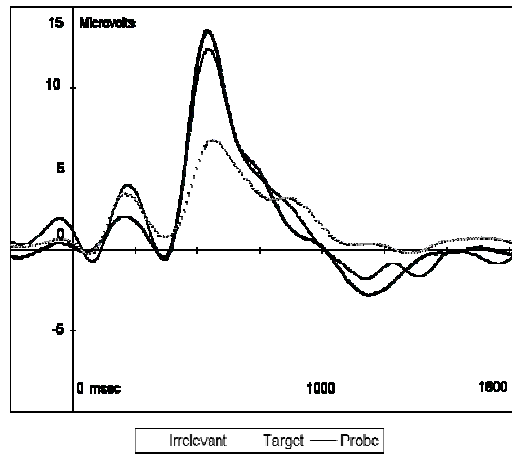

Figure 3 Information Present Brain Response

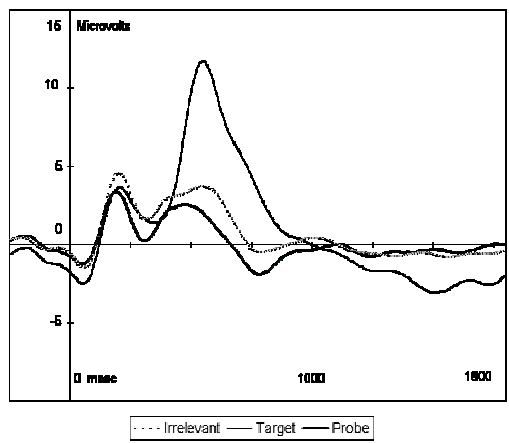

Figure 4 Information Absent Brain Response

The brain responses of two subjects indicates that the best stimulus to rely on in distinguishing between knowledgeable and not knowledgeable is the Probe signal.

## 5. Conclusions

Moving robots by mere thinking has been an awestruck approach for neural and muscular paralysis. The story of the Pied Piper inspired us to devise some new heuristics for interfacing human motor system using brain waves, by combining head helmet and LumbarMotionMonitor.

For the simulation we use java GridGain. Brain responses of classified subjects during training indicates that Probe can be the best stimulus to rely on in distinguishing between knowledgeable and not knowledgeable.

A long way is yet to go. Future work have many dimension. One dimension is to apply the heuristics for pervasive computing [24] . Other dimension is to enhance the accuracy of the system by mining event-related potential patterns[25] , or classifying the EEG data [26].